\documentclass[aip,jmp,graphicx,reprint]{revtex4-1}
\draft
\usepackage{dcolumn}
\usepackage{graphicx}
\usepackage{subfigure}
\usepackage{epstopdf}
\usepackage{color}

\newcommand*{\citen}[1]{%
  \begingroup
    \romannumeral-`\x 
    \setcitestyle{numbers}%
    \cite{#1}%
  \endgroup   
}
\begin{document}

\title{Temporal resolution criterion for correctly simulating relativistic electron motion in a high-intensity laser field}

\author{Alexey V. Arefiev}
\affiliation{Institute for Fusion Studies, The University of Texas, Austin, Texas 78712, USA}

\author{Ginevra E. Cochran}

\author{Douglass W. Schumacher}
\affiliation{Physics Department, The Ohio State University, Columbus, Ohio 43210, USA}

\author{\\Alexander P. L. Robinson}
\affiliation{Central Laser Facility, STFC Rutherford-Appleton Laboratory, Ditcot, OX11 0QX, UK}

\author{Guangye Chen}
\affiliation{Los Alamos National Laboratory, Los Alamos, New Mexico 87545, USA}

\date{\today}

\begin{abstract}

Particle-in-cell codes are now standard tools for studying ultra-intense laser-plasma interactions. Motivated by direct laser acceleration of electrons in sub-critical plasmas, we examine temporal resolution requirements that must be satisfied to accurately calculate electron dynamics in strong laser fields. Using the motion of a single electron in a perfect plane electromagnetic wave as a test problem, we show surprising deterioration of the numerical accuracy with increasing wave amplitude $a_0$ for a given time-step. We go on to show analytically that the time-step must be significantly less than $\lambda/c a_ 0$ to achieve good accuracy. We thus propose adaptive electron sub-cycling as an efficient remedy.

\end{abstract}

\maketitle

\section{Introduction}

Ongoing progress in laser engineering has significantly increased the maximum laser intensity available for ultra-intense laser-plasma experiments. A number of emerging applications now rely on the ability of high intensity laser beams to accelerate electrons to relativistic energies that considerably exceed the electron rest mass~\cite{Mangles2004,Geddes2004,Faure2004}. It has therefore become critical to accurately simulate electron dynamics in an ultra-intense electromagnetic field. However, it is not always possible to determine whether the simulation has been done with required accuracy and whether the simulation results are physically correct without performing a convergence study. Such a study can be a time-consuming effort without guarantee of a conclusive outcome if the parameter space has not been narrowed down sufficiently using relevant test cases. Convergence studies are best performed based on a fundamental understanding of the numerical and physical constraints on simulation parameters.

A commonly used tool for simulating laser-plasma interactions is a particle-in-cell (PIC) code that consists of two key blocks: a wave solver on a given spatial grid and a particle pusher that uses the calculated fields to advance particles. Two key parameters, besides the number of particles, that determine the accuracy and speed of the simulation are the grid-size and the time-step. The choice of the two is interrelated through the Courant criterion~\cite{Birdsall2002}, which limits the maximum time step for a given cell size, particularly for explicit codes. Depending on the problem, different criteria are used to determine the cell-size and the time-step. 

We are particularly interested in the regime where the plasma density is significantly below the critical density~\cite{Arefiev2014,Arefiev2012,Robinson2013} or, equivalently, where the laser frequency is significantly above the plasma frequency. In this regime, the group and phase velocities of an electromagnetic laser pulse inside the plasma are close to the speed of light, which enables electron acceleration to high energies as the electron moves forward with the pulse via direct laser acceleration. For this type of problem, one typically determines the spatial resolution first based on the wavelength of the laser pulse and its transverse dimensions. It is however not necessary to resolve the Debye length when simulating electron acceleration by an ultra-intense laser in a significantly under-dense plasma, since the energy that results from numerical heating is inconsequential (see Sec.~\ref{Sec_6} for a detailed explanation). The time-step is then determined using the Courant criterion, so that the numerical scheme remains stable. It is usually chosen close to its maximum value allowed by the Courant criterion in order to reduce the numerical dispersion of the wave caused by the grid. However, we find that at sufficiently high intensities in a low density environment, the wavelength does not set the scale for accurate treatment of electron motion. Instead, the electron motion near its stopping points becomes the critical factor resulting in surprisingly stringent requirements for convergence. Given the current widespread use of a large number of differing PIC codes, often incorporating multiple algorithms, simple test cases and criteria for evaluating them are highly beneficial. In particular, since experiments employing intensities of up to $10^{21}$ W/cm$^2$~[\citen{Bahk2004}] are now commonplace and experiments at significantly higher intensities are underway or anticipated~\cite{ELI}, evaluation of PIC codes in this regime is crucial.

We revisit the dynamics of a single free electron irradiated by a high-intensity plane electromagnetic wave as a test problem for evaluating the performance of a particle-in-cell code. The electron motion becomes relativistic at large normalized wave amplitudes, $a_0 \equiv |e| E / m_e c \omega \gg 1$, where $E$ is the amplitude of the wave electric field, $\omega$ is the wave frequency, $c$ is the speed of light,  and $e$ and $m_e$ are the electron charge and mass. In this regime, most of the electron energy is associated with the longitudinal motion and the maximum relativistic $\gamma$-factor increases as $\gamma \approx a_0^2/2$. As a result of the longitudinal motion with relativistic velocity, the frequency of the transverse electron oscillations decreases by a factor of $\gamma$, which can become substantial for large wave amplitudes. If, for a given time-step, the simulation reproduces the electron motion with $a_0 \sim 1$ correctly, one might expect that the relativistic motion of an electron with $a_0 \gg 1$ would also be correctly reproduced since the period of the oscillations increases with $a_0$. The numerical results presented below show an exactly opposite trend, with the accuracy quickly deteriorating with the increase of $a_0$ for a fixed time-step. 

In what follows, we outline a criterion that must be considered when simulating electron acceleration by a high amplitude electromagnetic wave in an under-dense plasma. We show that the electron dynamics can be correctly reproduced only if the time-step is sufficiently small to resolve the electron motion near stopping points along the trajectory. This condition requires that the time-step in the simulation is less than $1/a_0 \omega$, where $\omega$ is the wave frequency. This criterion is independent of constraints on spatial resolution. It becomes more stringent at higher wave amplitudes due to the fact that the acceleration is more rapid near the stopping points for larger $a_0$. This means that the error accumulates primarily along relatively small segments of the electron trajectory in the vicinity of the stopping points. We therefore propose adaptive electron sub-cycling as an efficient remedy. The idea is to reduce the time step for a given electron when the acceleration can no longer be correctly reproduced using the original time step. Our results show that sub-cycling permits a dramatic increase in accuracy with only a modest increase in the total number of time steps. Given current interest in direct laser acceleration~\cite{Gaillard2011,Kluge2012,Krygier2014}, the rapidly increasing focus on using ultra-intense lasers to study radiation reaction~\cite{Zhidkov2002,Ji2014} and the ambition to explore QED effects with lasers in the near future~\cite{Fedotov2010}, it is important that the new constraint on time-step described here be taken into account.

The rest of the paper is organized as follows. In Sec.~\ref{Sec_1}, we review the dynamics of a free electron irradiated by an incoming electromagnetic wave to establish the context for the analysis that follows. In Sec.~\ref{Sec_2}, we demonstrate using a significantly under-dense plasma that relative numerical errors grow with wave amplitude in the case of an electron accelerated by a laser pulse. In Secs.~\ref{Sec_3} and \ref{Sec_4} we analyze the errors originating from the particle pusher and derive a corresponding criterion for the time-step. 
In Sec.~\ref{Sec_5} we show that the errors can be greatly reduced using adaptive sub-cycling which helps to better resolve the electron dynamics near stopping points. Finally, in Sec.~\ref{Sec_6} we summarize our results and discuss possible implementation of the sub-cycling in a particle-in-cell code.


\section{Single electron dynamics in a plane wave} \label{Sec_1}

In this Section, we summarize the key features of single electron dynamics in a plane wave in order to establish the context for the subsequent analysis of the numerical results. We consider a free electron irradiated by a plane wave that propagates along the $x$-axis. The wave electric field is directed along the $y$-axis and the wave magnetic field is directed along the $z$-axis. The wave propagation can be described using a normalized vector potential
\begin{equation}
{\bf{a}} (x,t) = a(\xi) {\bf{e}}_y, 
\end{equation}
where $a$ is only a function of a dimensionless phase variable
\begin{equation} \label{Eq_xi}
	\xi \equiv 2 \pi (ct - x) / \lambda.
\end{equation}
Here $\lambda$ is the wave-length, $c$ is the speed of light, $t$ is the time in the laboratory frame of reference, and ${\bf{e}}_y$ is a unit vector. The electric and magnetic fields of the wave are given by
\begin{eqnarray}
&& E = - \frac{m_e c}{|e|} \frac{\partial a}{\partial t}, \label{Eq_E}\\
&& B = \frac{m_e c^2}{|e|} \frac{\partial a}{\partial x}, \label{Eq_B}
\end{eqnarray}
where $m_e$ and $e$ are the electron mass and charge.

An initially stationary electron irradiated by this wave moves only in the $(x,y)$-plane according to the following equations: 
\begin{eqnarray}
&& \frac{d}{d t} \left( \frac{p_x}{m_e c} \right)  = - \frac{|e|B}{\gamma m_e c} \frac{p_y}{m_e c}, \label{Eq2.1}\\
&& \frac{d}{d t} \left( \frac{p_y}{m_e c} \right)  = -\frac{|e| E}{m_e c} + \frac{|e|B}{\gamma m_e c} \frac{p_x}{m_e c} , \label{Eq2}\\
&& \frac{dx}{d t} =  \frac{c}{\gamma}\frac{p_x}{m_e c}, \label{Eq4} \\
&& \frac{dy}{d t} =  \frac{c}{\gamma}\frac{p_y}{m_e c}, \label{Eq4.2}
\end{eqnarray}
where $p_x$ and $p_y$ are components of the electron momentum and
\begin{equation}
\gamma = \sqrt{1 + \left(p_x / m_e c \right)^2 + \left(p_y / m_e c \right)^2} \label{Eq5}
\end{equation}
is the relativistic factor. This system of equations has two integrals of motion:
\begin{eqnarray}
&& \frac{d}{dt} \left( \frac{p_y}{m_e c} - a \right) = 0, \label{Integral_1} \\
&& \frac{d}{dt} \left( \gamma - \frac{p_x}{m_e c} \right) = 0. \label{Integral_2} 
\end{eqnarray}
We skip the derivation here, which can, for example, be found in the references~[\citen{Boyd2003}] and [\citen{Arefiev2014}].

The second integral of motion implies that the electron dephases from the wave at a constant rate. In order to show that, we first take the derivative of the phase variable $\xi$ defined by Eq.~(\ref{Eq_xi}) with respect to time $t$, which yields 
\begin{equation} \label{deph_1}
\frac{d \xi}{d t} = \frac{\omega}{\gamma} \left(\gamma - \frac{p_x}{m_e c} \right).
\end{equation}
On the other hand, the proper time that we denote as $\tau$ and the time $t$ are related by the expression $d \tau / d t = 1/\gamma$. The proper time is the elapsed time that would be measured by the electron itself. Using this relation in Eq.~(\ref{deph_1}), we find that 
\begin{equation} \label{deph_2}
\frac{d \xi}{d \tau} = \omega \left(\gamma - \frac{p_x}{m_e c} \right).
\end{equation}
According to the integral of motion (\ref{Integral_2}), the expression on the right-hand side is a constant and, therefore, $d \xi / d \tau$ is also a constant. This means that the phase of the field sampled by the electron increases linearly with proper time. It is then appropriate to interpret the integral of motion (\ref{Integral_2}) as the corresponding dephasing rate. It should be emphasized that the value of the dephasing rate has a direct and significant impact on the maximum energy that the electron gains during acceleration by the wave~\cite{Arefiev2014,Robinson2013}. This aspect will play a key role in the subsequent analysis in Secs. \ref{Sec_2} - \ref{Sec_5}.

If the electron is at rest ($p_x = p_y = 0$) before the wave arrives ($a = 0$), then it follows from Eq.~(\ref{Integral_2}) that 
\begin{equation} \label{Integral_2_2}
\gamma - \frac{p_x}{m_e c} = 1.
\end{equation}
Using Eqs.~(\ref{Integral_1}), (\ref{Integral_2_2}), and the definition of $\gamma$, we find that
\begin{eqnarray}
&& p_y / m_e c = a, \label{Eq_py}\\
&& p_x / m_e c = a^2/2. \label{Eq_px}
\end{eqnarray}
In momentum space, the electron always moves along a parabola $p_x/m_e c = (p_y/m_e c)^2/2$, with only the maximum displacement along $p_x$ and $p_y$ changing with the wave amplitude. 

Equations (\ref{Eq_py}) and (\ref{Eq_px}) give $p_x$ and $p_y$ only in terms of $\xi$ and one still needs to integrate Eq.~(\ref{Eq4}) in order to find $p_x(x,t)$ and $p_y(x,t)$. One can find from Eq.~(\ref{Eq4}) using the definition of $\xi$ given by Eq.~(\ref{Eq_xi}) that
\begin{eqnarray}
&& \frac{ct}{\lambda} = \frac{1}{2 \pi} \int_0^{\xi} \gamma d \xi', \label{Eq_t}\\
&& \frac{x}{\lambda} = \frac{1}{2 \pi} \int_0^{\xi} \left( \gamma - 1 \right) d \xi' \label{Eq_x},
\end{eqnarray}
where $\gamma = 1 + a^2(\xi) / 2$ according to Eqs.~(\ref{Eq5}), (\ref{Eq_py}), and (\ref{Eq_px}). Equations (\ref{Eq_py}) - (\ref{Eq_x}) allow one to implicitly determine components of the electron momentum as functions of time $t$ and axial distance $x$.

To summarize, a free electron irradiated by a plane wave moves along a parabola in momentum space due to the fact that its dephasing rate $\gamma - p_x/m_e c$ remains constant. The maximum electron $\gamma$-factor is $\gamma_* = 1 + a_0^2/2$ and the corresponding maximum electron energy gain is $\gamma_* m_e c^2$, where $a_0$ is the maximum value of $a$. 


\section{Simulation results for a significantly under-dense plasma} \label{Sec_2}

In this Section, we present several results from particle-in-cell simulations in order to determine how well the single electron dynamics described in Sec.~\ref{Sec_1} is reproduced numerically. We initialize a low-density hydrogenic plasma slab ($n_e = n_i = 10^{-3}$ $n_{\mbox{crit}}$, where $n_{\mbox{crit}}$ is the critical density) with cold electrons that are irradiated by a plane electromagnetic wave. The plasma density is deliberately set very low so that space-charge effects are negligible during our runs and each electron effectively behaves as a free electron in a vacuum. The plasma is essentially acting as a convenient cold electron source. At this density, the effect of the electron currents on the field of the wave is also negligible in our runs. This setup mimics the initial conditions considered in Sec.~\ref{Sec_1}. The normalized pulse amplitude $a$ ramps up to $a_0$ and then remains constant. The exact profile of the electron density and the wave amplitude are not important. All the plasma electrons are equivalent in this setup and they only differ by their initial location. In what follows, we select one electron and track it throughout the simulation.

\begin{figure}[tb]
  \centering
  \subfigure{\includegraphics[scale=0.4]{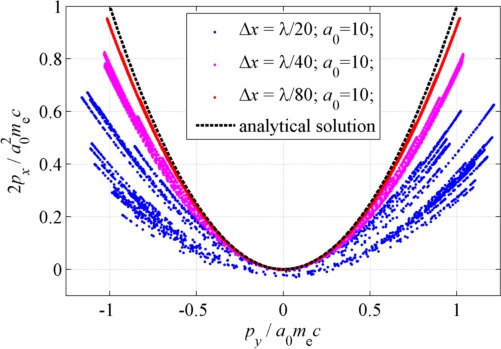} } \\
  \subfigure{\includegraphics[scale=0.4]{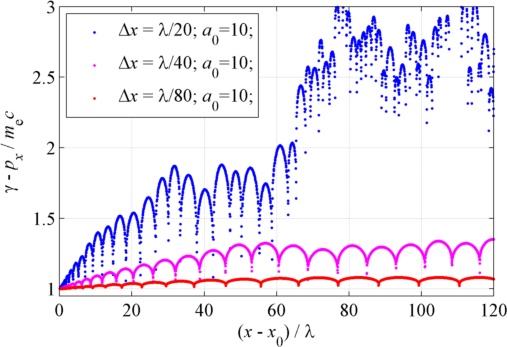} }
\caption{Electron momentum space (top) and the dephasing rate (bottom) for different values of grid-size, $\Delta x$, at $a_0 = 10$. For all three runs, we set $c \Delta t / \Delta x = 0.95$.}
\label{fig:phase_space_a0_10}
\end{figure}

Figure \ref{fig:phase_space_a0_10} shows numerical results for $a_0 = 10$ and three different grid sizes, $\Delta x/\lambda = 1/20$, 1/40, and 1/80. In all three runs, the ratio of the time-step $\Delta t$ to the grid-size $\Delta x$ was the same, with $c \Delta t / \Delta x = 0.95$. The upper panel in Fig. \ref{fig:phase_space_a0_10} shows the trajectory of a single electron in momentum space and the lower panel shows the corresponding dephasing rate as a function of the distance traveled by the electron from its initial location. The electron data is shown with dots, because it was recorded at discrete time intervals $dt = 0.075 \lambda / c$. According to the analytical solution of Sec.~\ref{Sec_1}, the electron should be moving along a parabola $p_x/m_e c = (p_y/m_e c)^2/2$, with $|p_y/m_e c| \leq a_0$ and $0 \leq p_x/m_e c \leq a_0^2/2$. The dephasing rate must remain constant and equal to unity, $\gamma - p_x/m_e c = 1$. Not surprisingly, the convergence to the analytical solution improves as we decrease $\Delta x$, which in this case is equivalent to decreasing $\Delta t$ because their ratio is maintained. 

\begin{figure}[tb]
  \centering
  \subfigure{\includegraphics[scale=0.4]{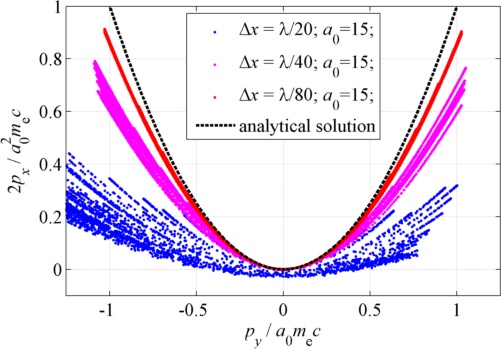} } \\
  \subfigure{\includegraphics[scale=0.4]{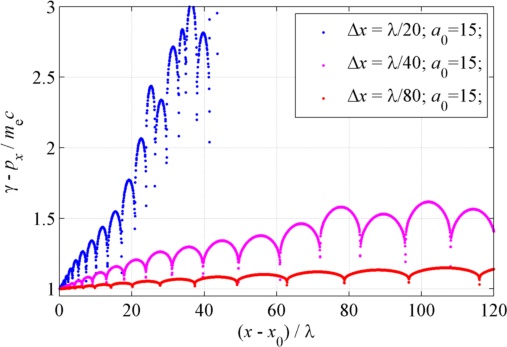} }
\caption{Electron momentum space (top) and the dephasing rate (bottom) for different values of grid-size, $\Delta x$, at $a_0 = 15$. For all three runs, we set $c \Delta t / \Delta x = 0.95$.}
\label{fig:phase_space_a0_15}
\end{figure}

Figure~ \ref{fig:phase_space_a0_10} shows that there is a correlation between the deviation of the dephasing rate from unity and the deviation of the numerical solution in momentum space from the analytical result. This trend is also not surprising, because the dephasing rate determines how the field acting on the electron changes in time. An error in the dephasing causes an error in the electron acceleration and, consequently, leads to an error in the electron momentum. Note that errors in the dephasing rate also lead to considerable asymmetry in the transverse momentum $p_y$ (see the results for $\Delta x = \lambda/20$ in Figs.~ \ref{fig:phase_space_a0_10} and \ref{fig:phase_space_a0_15}), which can result in an unphysical electron drift perpendicular to the direction of the wave propagation. 

There is a distinct periodic structure of sharp downwards spikes in the dephasing rate. By comparing the time evolution of the dephasing rate and the electron momentum, we find that the downwards spikes correspond to stopping points. The numerical errors in the dephasing rate change considerably along the electron trajectory and they are the most significant in the vicinities of the stopping points. This is immediately evident from the change in the vertical distance between adjacent data points (dots) in the lower panel of Fig.~ \ref{fig:phase_space_a0_10}. As stated earlier, the time interval between the adjacent data points is constant and, therefore, the dephasing rate changes at a greater rate around the downwards spikes.

Figure~\ref{fig:phase_space_a0_15} shows the same quantities as Fig.~ \ref{fig:phase_space_a0_10}, but for a higher wave amplitude of $a_0 = 15$. The already discussed trends seem to be similar in this case. We deliberately used the same set of grid sizes and the same ratio $c \Delta t / \Delta x$ as in the case of $a_0 = 10$ in order to determine how numerical errors scale with wave amplitude. Let us examine the runs with $\Delta x = \lambda/40$ for $a_0=10$ and $a_0=15$. The period of electron oscillations in the laser field increases with $a_0$ due to the increased $\gamma$-factor that is primarily associated with the longitudinal motion. This can be seen by comparing the distance between the downwards spikes in the dephasing rate that is also the distance between stopping points. Since the velocity of the longitudinal motion is close to $c$, longer distance translates directly into a longer interval between stopping points and thus a longer period of oscillations. The time step for both runs at  $a_0=10$ and $a_0=15$ is the same, so one might expect that the higher amplitude run would be better resolved and the numerical errors would be reduced. The comparison of the deviation in the dephasing rate from unity indicates that the trend is exactly the opposite. Greater errors in the dephasing rate then lead to greater errors in the electron momentum at higher wave amplitude. 

We therefore conclude that the numerical accuracy deteriorates with increasing wave amplitude for a fixed time step despite the fact that the period of the electron oscillations increases. In the following Sections we examine the source of the discovered increasing inaccuracy.


\section{Analysis of errors originating from the particle-pusher} \label{Sec_3}

In general, there is a wide range of factors that contribute to numerical errors in a particle-in-cell simulation~(see reference [\citen{Godfrey2013}], [\citen{Vay2008}], [\citen{Esirkepov2001}], [\citen{Birdsall2002b}] and references therein). The results of Sec.~\ref{Sec_2} indicate that in our test problem the numerical errors in the dephasing rate tend to significantly increase around specific points of the electron trajectory - the stopping points. This observation serves as a motivation for us to consider the particle pusher separately. 

In what follows, we specify the wave field analytically. In the test problem under consideration, this is simply an electromagnetic pulse propagating in a vacuum. This approach allows us to isolate the errors introduced by the particle pusher. We use the standard Boris pusher \cite{Birdsall2002} to advance the electron momentum and coordinates in time using the analytical solution for the field at the electron location. The following results are for an initially immobile single electron irradiated by a plane electromagnetic wave.

\begin{figure}[tb]
  \centering
  \subfigure{\includegraphics[scale=0.4]{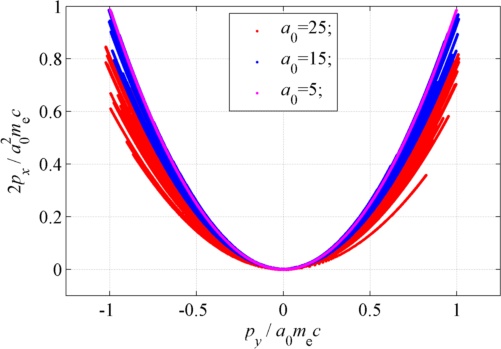} } \\
  \subfigure{\includegraphics[scale=0.4]{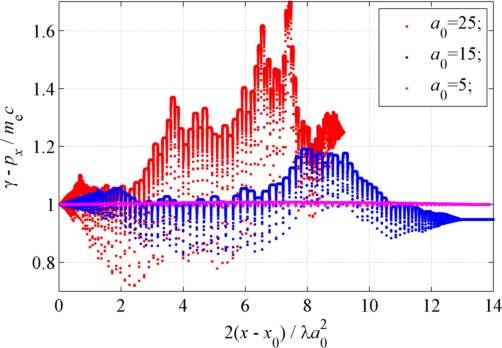} }
\caption{Electron momentum space (top) and the dephasing rate (bottom) for $a_0 = 5$, 15, and 25 calculated using a particle pusher and an analytical field. We set $c \Delta t / \lambda = 1/50$ for all three runs.}
\label{fig:pusher_0}
\end{figure}

Figure~\ref{fig:pusher_0} shows the results for three different wave amplitudes and the same time-step, $c \Delta t / \lambda = 1/50$. The upper panel shows the electron momentum space. Deviation from the analytical solution predicting $p_x/m_e c = (p_y/m_e c)^2/2$ [see Eqs.~(\ref{Eq_py}) and ~(\ref{Eq_px}) ] increases with wave amplitude. The lower panel shows the dephasing rate as a function of the longitudinal displacement. The errors in the dephasing rate and the deviation from $\gamma - p_x/m_e c = 1$ also increase with $a_0$. These trends are similar to those observed in Sec.~\ref{Sec_2} where the wave fields were calculated numerically using a finite difference scheme for the Maxwell equations. This suggests that the increase in numerical errors is caused by the particle pusher, whereas the errors resulting from numerical integration of the field equations are less significant for the grid-size and time-step used.

\begin{figure}[tb]
  \centering
  \subfigure{\includegraphics[scale=0.4]{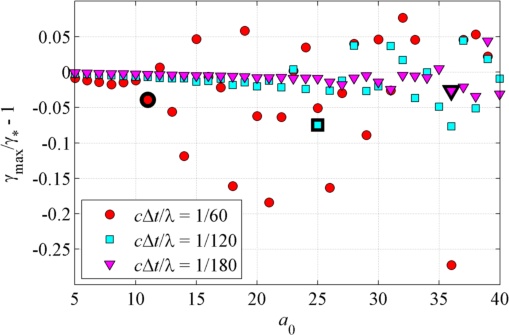} } \\
  \subfigure{\includegraphics[scale=0.4]{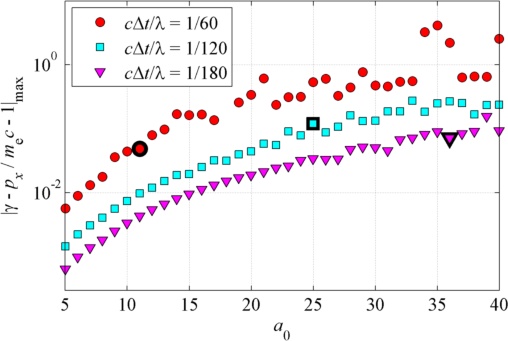} }
\caption{Field amplitude scans of the relative error in the electron energy gain (top) and the error in the dephasing rate (bottom) using three different time steps: $c \Delta t / \lambda = 1/60$, 1/120, and 1/180. See discussion in text of these figures of merit. The enlarged markers indicate the threshold intensity for the onset of large errors in the electron energy gain for each value of $c \Delta t / \lambda$ (see Table~\ref{table}).}
\label{fig:pusher_00}
\end{figure}

In order to see the trend for the numerical errors more clearly, we have performed field amplitude scans for three different time-steps, $c \Delta t / \lambda = 1/60$, 1/120, and 1/180. The results are shown in Fig.~\ref{fig:pusher_00} for $5 \leq a_0 \leq 40$. The upper panel shows the discrepancy in the electron energy gain and the lower panel shows the discrepancy in the dephasing rate. For each field amplitude, we determine the maximum $\gamma$-factor achieved by the electron in the simulation, denoted as $\gamma_{\mbox{max}}$. The analytical solution in Sec.~\ref{Sec_1} predicts the maximum $\gamma$-factor to be
\begin{equation}
\gamma_* = 1 + a_0^2/2,
\end{equation}
where $a_0$ is the peak amplitude of the pulse. The quantity plotted in the upper panel of Fig.~\ref{fig:pusher_00} is $(\gamma_{\mbox{max}} - \gamma_*)/\gamma_*$, which is the relative error in the predicted electron energy gain.  For each wave amplitude, we also determine the maximum absolute deviation of the dephasing rate $\gamma - p_x/m_ec$ from unity. This is the quantity plotted in the lower panel of Fig.~\ref{fig:pusher_00}.

There are several important trends that become apparent from the scans presented in Fig.~\ref{fig:pusher_00}. The error in the dephasing rate gradually increases with $a_0$ regardless of the time-step used to integrate the electron equations of motion. In contrast with that, the discrepancy in the electron energy gain exhibits a threshold behavior with the increase of $a_0$. The threshold for the onset of large errors in the electron energy gain appears to scale inversely proportional to the time-step used in the particle pusher. If, for example, we define this threshold as $|\gamma_{\mbox{max}} - \gamma_* |/\gamma_* = 0.025$, then we find that the discrepancy exceeds the threshold value at $a_0 = 11$, 25, and 36 for $c \Delta t / \lambda = 1/60$, 1/120, and 1/180. This is indeed a threshold, since the discrepancy in the energy gain jumps up considerably at these wave amplitudes. We have listed these numbers in Table~\ref{table}. We have also listed in Table~\ref{table} $|\gamma - p_x/m_ec - 1|_{\mbox{max}}$ above the threshold. The threshold clearly occurs at roughly the same level of discrepancy in the dephasing rate in all three cases. The values from Table~\ref{table} are shown with enlarged markers in Fig.~\ref{fig:pusher_00}.

\begin{table}[ht]
\caption{Threshold for errors in the energy gain} 
\centering 
\begin{tabular}{c c r} 
\hline\hline 
$c \Delta t / \lambda$ & $a_0$   & $|\gamma - p_x/m_ec - 1|_{\mbox{max}}$ \\ [0.5ex] 
\hline 
1/60 & 11 & 0.080 \\ 
1/120 & 25 & 0.088 \\
1/180 & 36 & 0.083 \\ [1ex] 
\hline 
\end{tabular}
\label{table} 
\end{table}

We can summarize this Section by concluding that there is a general trend for the errors in the dephasing rate to increase with wave amplitude. Once the discrepancy in the dephasing rate approaches 10\%, a rapid increase in the discrepancy in the electron energy gain takes place. The corresponding threshold wave amplitude scales inversely proportional to the time-step used to integrate the electron equations of motion.


\section{Criterion for the time-step} \label{Sec_4}

In the previous Sections, we showed that the numerical errors tend to increase with wave amplitude. In this Section, we formulate a criterion for the time-step that must be satisfied in order to accurately reproduce the electron dynamics in a strong electromagnetic wave.

It is helpful to begin by reviewing how a standard Boris particle pusher advances the electron momentum in time. This is done in three subsequent stages using given electric and magnetic fields. It first accelerates the electron using the electric field. It then performs a rotation in the magnetic field. The last stage is another push using the electric field. The (half) rotation angle is 
\begin{equation} \label{rot_angle}
\psi = - \frac{1}{2}\frac{|e| B \lambda}{\gamma m_e c^2} \frac{c \Delta t}{\lambda},
\end{equation}
where $B$ is a given magnetic field and $\gamma$ is the relativistic factor of the electron after the first push by the electric field. 

This procedure necessarily requires the rotation angle to be small, which can be easily understood in the context of the test problem that we are considering. As the electron is pushed forward by the wave, its transverse momentum oscillates, while the longitudinal momentum remains positive. Therefore, the electron never performs a full rotation in momentum space. On the other hand, if the rotation angle in the Boris pusher is comparable to $\pi$, then the particle pusher qualitatively changes the electron motion causing the electron to move backwards.   

Our next step is to determine the relation between the rotation angle and the wave amplitude in our test problem.
Let us take a pulse that, after some initial ramp-up, has a constant amplitude, with $a = a_0 \sin(\xi)$. In this case,  the electric and magnetic fields acting on the electron are
\begin{equation} \label{fields}
E = B = - m_e c^2 a_0 \cos(\xi) 2 \pi / \lambda |e|.
\end{equation}
According to the analytical solution of Sec.~\ref{Sec_1}, the $\gamma$-factor of the electron is
\begin{equation}
\gamma = 1 + a^2/2.
\end{equation}
The ratio $B/\gamma$ has the largest absolute value for $a = 0$. Thus the rotation angle for a given time-step $\Delta t$ has also the largest value for $a = 0$,
\begin{equation} \label{rot_angle_2}
{\mbox{max}} |\psi| = \pi a_0 \frac{c \Delta t}{\lambda}.
\end{equation}
The requirement max$|\psi| \ll \pi$ now yields
\begin{equation} \label{main_criterion}
\frac{c \Delta t}{\lambda} \ll \frac{1}{a_0}.
\end{equation}
This condition indicates that the wave amplitude imposes an upper limit on the time-step that can be used by the particle pusher. It should be noted that the discussed criterion is equivalent to the requirement that the smallest time scale in the problem must be resolved. In our case, the restriction is imposed by the wave magnetic field and the corresponding time scale that must be resolved is the gyro-period at $a=0$.

The points along the electron trajectory where $a$ vanishes are stopping points. This is evident from the analytical solution of Sec.~\ref{Sec_1}. Equations~(\ref{Eq_py}) and (\ref{Eq_px}) give $p_y$ and $p_z$ as functions of $a$, with both vanishing for $a=0$. On the other hand, the electric and magnetic fields of the wave reach their maximum amplitude at a stopping point. Therefore, the vicinity of a stopping point is that part of the electron trajectory where the electron experiences the strongest acceleration and, as a result, the rotation angle has the largest amplitude. 

It might seem that the derived criterion for the time-step is unnecessarily restrictive due to the fact that it is imposed by electron dynamics near stopping points. Indeed, even if the criterion is not satisfied, the electron momentum gain that would be calculated incorrectly would still be relatively small compared to both $|p_y| = a_0$ and $p_x = a_0^2/2$. However, this argument does not take into account the corresponding error in the dephasing rate and its impact on the subsequent electron acceleration by the wave.

In order to show that errors in acceleration near stopping points are indeed critical, let us first estimate how the dephasing rate changes as a result of these errors. We define the dephasing rate as
\begin{equation} \label{deph_est}
I \equiv \gamma - \frac{p_x}{m_e c} = \sqrt{1 + \left( \frac{p_x}{m_e c} \right)^2 + \left( \frac{p_y}{m_e c} \right)^2} - \frac{p_x}{m_e c}.
\end{equation} 
Let us consider an electron as it starts its motion right at the stopping point with correct initial conditions determined from the analytical solution, so that $p_x = p_y = 0$ and $a = 0$. The electron dephasing rate is then $I = 1$. We now use a standard Boris pusher to advance the electron momentum by one time-step. We assume that the rotation is calculated incorrectly and we want to estimate the resulting error in the dephasing rate. The momentum gain resulting from the acceleration by the wave electric field over a time interval equal to $\Delta t$ is roughly $|\Delta p / m_e c| = 2 \pi a_0 c \Delta t/ \lambda$. Deliberately assuming the worst case scenario, we set $p_y = 0$ and $p_x = \pm |\Delta p|$ in Eq.~(\ref{deph_est}). For $|\Delta p / m_e c| \ll 1$, we have the following estimate for the dephasing after just one time-step: 
\begin{equation} \label{deph_error}
I \approx 1 \pm |\Delta p / m_e c| \approx 1 \pm 2 \pi a_0 c \Delta t/ \lambda. 
\end{equation}

An electron with an initial axial momentum $p_x = \pm |\Delta p|$ would have the same dephasing rate. An analytical solution for such an electron is given in Ref.~[\citen{Arefiev2014}]. The maximum electron $\gamma$-factor according to Eq.~(25) of Ref.~[\citen{Arefiev2014}] is $\gamma_{\mbox{max}} = (1+ a_0^2 + I^2)/ 2I$. The difference in the maximum $\gamma$-factor achieved by this electron as compared to the case considered in Sec.~\ref{Sec_1} where the electron is initially at rest is
\begin{equation} \label{gain_error}
\frac{\gamma_{\mbox{max}} - \gamma_*}{\gamma_*} = \left(1 - I \right) \frac{1 + a_0^2 - I}{I \left( 2 + a_0^2 \right)}.
\end{equation}
We can now employ this expression to estimate the impact that an error in the dephasing has on the electron energy gain. We use the estimate for the dephasing rate given by Eq.~(\ref{deph_error}) and assume that the error in the dephasing rate is relatively small, $|1 - I| \ll 1$. It then follows from Eq.~(\ref{gain_error}) that 
\begin{equation} \label{gain_error_2}
\frac{\gamma_{\mbox{max}} - \gamma_*}{\gamma_*} \approx 1 - I \approx \pm 2 \pi a_0 \frac{c \Delta t}{\lambda}.
\end{equation}
This estimate indicates that the relative error in the electron energy gain is of the same order as the error in dephasing rate.  

These estimates elucidate the physical basis for the restriction on the time-step given by Eq.~(\ref{main_criterion}). If the condition~(\ref{main_criterion}) is not satisfied, then the numerical error resulting from numerical integration of the electron equations of motion near a stopping point leads to a considerable error in the dephasing rate. The error in the momentum gain is relatively small at this stage. However, the error in the dephasing rate affects the subsequent electron acceleration by the wave even if no additional errors are introduced, causing a considerable error in the maximum electron energy gain.


\section{Adaptive sub-cycling} \label{Sec_5}

We have so far examined the errors introduced by a standard Boris pusher and determined that the electron dynamics in the vicinity of stopping points imposes a stringent upper limit on the time-step. Guided by these observations, we develop in this Section a procedure that allows us to considerably improve the accuracy when calculating the electron dynamics with only a modest increase of the total number of time steps.

\begin{figure}[tb]
  \centering
  \subfigure{\includegraphics[scale=0.4]{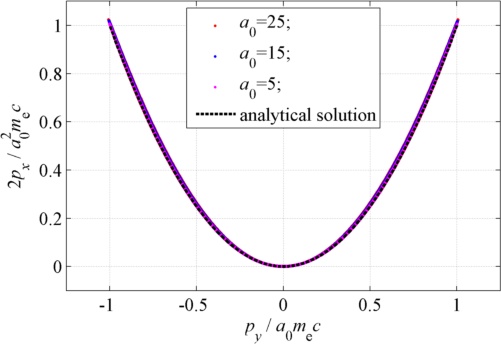} } \\
  \subfigure{\includegraphics[scale=0.4]{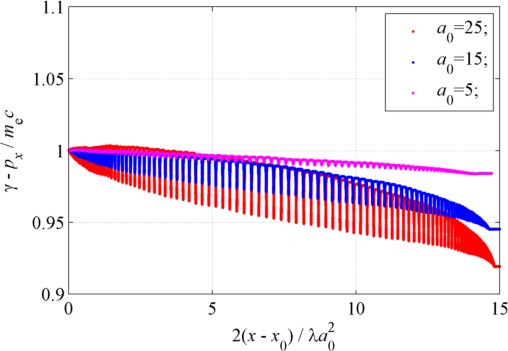} }
\caption{Electron momentum space (top) and the dephasing rate (bottom) for $a_0 = 5$, 15, and 25 calculated using adaptive sub-cycling. The particle pusher uses an analytical field. The base time-step is set at $c \Delta_0 / \lambda = 1/50$ for all three runs.}
\label{fig:pusher_0s}
\end{figure}

A direct way to reduce numerical errors generated by the particle pusher is by decreasing the time-step $\Delta t$. However, if the time-step reduction necessary to satisfy the criterion (\ref{main_criterion}) is significant, then this approach would greatly increase the total number of time-steps needed to simulate the same time interval. On the other hand, the increased precision is not helpful and thus unnecessary for the majority of the electron trajectory.

A more efficient way to reduce the numerical errors is by adaptively decreasing the time-step only when necessary to ensure that the criterion (\ref{main_criterion}) is always satisfied. We implement this by introducing a 
critical rotation angle $\psi_*$. We also choose a base value for the time-step $\Delta t = \Delta_0$. At the beginning of each time-step, we estimate the rotation angle for the particle pusher, $\psi_{est}$, using Eq.~(\ref{rot_angle}), where the value of $\gamma$ is taken from the previous time-step and $\Delta t = \Delta_0$. If this estimate exceeds $\psi_*$, then we reduce $\Delta t$, which reduces $\psi_{est}$, to make $\psi_{est}$ less than $\psi_*$. For convenience, we choose the reduced time-step from a list of discrete values $\Delta t = \Delta_0/4^k$, where $k = 1,2,3,...$.  We pick the largest value that yields $\psi_{est} < \psi_*$. The algorithm automatically resets $\Delta t$ to the base value $\Delta_0$ if a reduced time-step is no longer necessary.

The standard Boris pusher updates electron position and momentum at interleaved time points, staggered such that they 'leapfrog' over each other. In order to successfully implement the described algorithm, one must synchronize the electron position and momentum at a time point corresponding to the momentum before changing the time-step. Therefore, the particle pushing algorithm should involve the following two steps after the momentum has been updated and the time step has to be changed. The first step is to advance the electron position by $\Delta t/2$, where $\Delta t$ is the original time-step. The second step is to advance the electron position by $\Delta t/2$, where $\Delta t$ is the new time-step. After this, the standard particle pushing algorithm can be used with a new time-step, starting with a momentum update. 

\subsection{Sub-cycling example}

As an example, we re-run the three cases presented in Fig.~\ref{fig:pusher_0} (see Sec.~\ref{Sec_3}) now using adaptive sub-cycling. The base time-step for all three runs is set to $c \Delta_0 / \lambda = 1/50$, which was the time-step used to generate Fig.~\ref{fig:pusher_0}. The critical rotation angle is set to $\psi_* = 0.05$. The resulting electron momentum and electron dephasing are shown in Fig.~\ref{fig:pusher_0s}. The deviation of the dephasing rate from unity has been significantly reduced, as compared to the results in the lower panel in Fig.~\ref{fig:pusher_0}. It is less than 10\% even for $a_0 = 25$. As a result, the deviation of the electron momentum from the analytical solution given in Sec.~\ref{Sec_3} is no longer visually detectable. It has been dramatically reduced compared to the results in the upper panel in Fig.~\ref{fig:pusher_0}. 

In Table \ref{table2}, we have listed how many times each time-step value is used during the sub-cycling. The relative number of steps using the base time-step value increases with increasing $a_0$, so that the number of reduced time-steps is less than 13\% for $a_0 = 25$. Recall that the same value of $a_0$ had the largest deviation from the analytical solution in Fig.~\ref{fig:pusher_0}. Therefore, the sub-cycling algorithm becomes more efficient at higher wave amplitudes and a significant improvement of the numerical results can be achieved using only a modest increase in the total number of time-steps. This is not surprising, since the electron spends only a small fraction of its time near the stopping points, as the time interval between the stopping points increases as $\gamma_{\max} \propto a_0^2$.

\begin{table}[ht]
\caption{Number of time-steps used for sub-cycling} 
\centering 
\begin{tabular}{c c c c c c} 
\hline\hline 
$a_0$ & $\psi_*$ & $\Delta t$  & $\Delta t/4$ & $\Delta t/16$ & $\Delta t/64$ \\ [0.5ex] 
\hline 
5 & 0.05 & 58\% & 30\% & 12\% & 0\% \\ 
15 & 0.05 & 78\% & 13\% & 8\% & 1\% \\
25 & 0.05 & 87\% & 7\% & 4\% & 2\% \\ [1ex] 
\hline 
\end{tabular}
\label{table2} 
\end{table}

\begin{figure}[tb]
  \centering
  \subfigure{\includegraphics[scale=0.4]{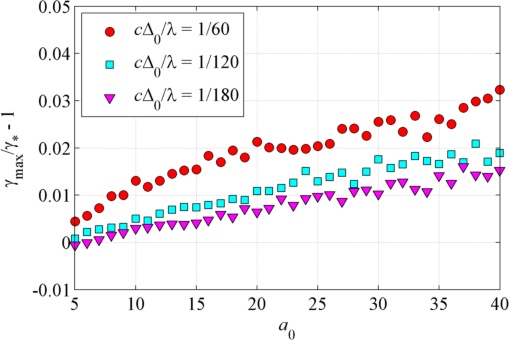} } \\
  \subfigure{\includegraphics[scale=0.4]{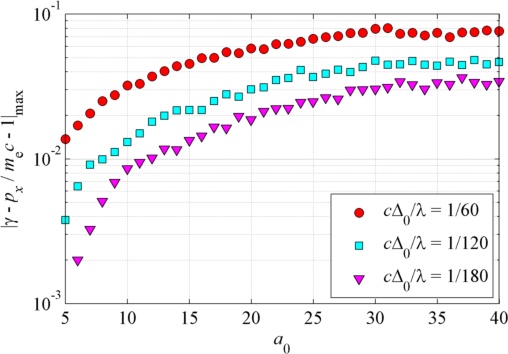} }
\caption{Field amplitude scans of the relative error in the electron energy gain (top) and the error in the dephasing rate (bottom) using adaptive sub-cycling for base time-step values $c \Delta_0 / \lambda = 1/60$, 1/120, and 1/180.}
\label{fig:pusher_00s}
\end{figure}

We have performed wave amplitude scans for three different base time-step values, $c \Delta_0 / \lambda = 1/60$, 1/120, and 1/180 for $5 \leq a_0 \leq 40$. The critical rotation angle was set to $\psi_* = 0.05$. Without sub-cycling, these three scans would produce the results shown in Fig.~\ref{fig:pusher_00}. The results using sub-cycling are shown in Fig.~\ref{fig:pusher_00s}. As in the case of Fig.~\ref{fig:pusher_00}, the upper panel shows the discrepancy in the electron energy gain and the lower panel shows the discrepancy in the dephasing rate. For each wave amplitude, we determine the maximum $\gamma$-factor achieved by the electron in the simulation, denoted as $\gamma_{\mbox{max}}$. The analytical solution in Sec.~\ref{Sec_1} predicts the maximum $\gamma$-factor to be $\gamma_* = 1 + a_0^2/2$, where $a_0$ is the peak amplitude of the pulse. The quantity plotted in the upper panel of Fig.~\ref{fig:pusher_00s} is $(\gamma_{\mbox{max}} - \gamma_*)/\gamma_*$, which is the relative error in the predicted electron energy gain.  For each wave amplitude, we also determine the maximum absolute deviation of the dephasing rate $\gamma - p_x/m_ec$ from unity. This is the quantity plotted in the lower panel of Fig.~\ref{fig:pusher_00s}.

The trends seen in Fig.~\ref{fig:pusher_0s} are confirmed by the wave amplitude scans in Fig.~\ref{fig:pusher_00s}. The deviation from the analytical solution of the dephasing and the electron energy gain have been dramatically reduced. The errors in the dephasing rate remain below 10\% for all runs. The errors in the electron energy gain now increase gradually with $a_0$ and do not exhibit the threshold behavior seen in Fig.~\ref{fig:pusher_00}. This is due to the fact that the amplitude of the rotation angle in the particle pusher always remains smaller than $\psi_* = 0.05$, which prevents the particle pusher from introducing qualitative changes to the electron dynamics near the turning points.

\subsection{Efficiency of the sub-cycling algorithm}

The key metric of the proposed algorithm is the relative increase in the total number of time steps between two stopping points due to the sub-cycling. In order to evaluate this quantity, we consider a vicinity of a stopping point $\xi=0$ where the wave amplitude is $a(\xi) = a_0 \sin(\xi)$. We assume that $a_0 \gg 1$ and that the condition given by Eq.~(\ref{main_criterion}) is not satisfied for $\Delta t = \Delta_0$, so that the sub-cycling is required. 

The fact that the criterion (\ref{main_criterion}) is not satisfied implies that the rotation angle $\psi$ defined by Eq.~(\ref{rot_angle}) exceeds $\psi_*$ at $\xi=0$. As the electron accelerates from the stopping point, the rotation angle decreases with the increase of $\xi$. We use Eq.~(\ref{rot_angle}) together with the expressions for the fields given by Eq.~(\ref{fields}) to find that
\begin{equation} \label{psi_1}
\psi(\xi) = \pi \frac{c \Delta t}{\lambda} \frac{a_0 \cos(\xi)}{1+a_0^2 \sin^2(\xi) / 2}. 
\end{equation}
The condition $\psi(0) = \psi_*$ sets the smallest time-step $\Delta t_{\min}$ that the sub-cycling procedure would have to use, which is
\begin{equation}
\frac{c \Delta t_{\min}}{\lambda} = \frac{\psi_*}{\pi a_0}.
\end{equation} 

The sub-cycling is only needed for $|\xi| < \xi_*$, where $\xi_*$ is defined by the condition $\psi(\xi_*) = \psi_*$. Away from the stopping point, where $\psi < \psi_*$, the sub-cycling is not necessary. We assume that $\psi$ matches $\psi_*$ close to $\xi = 0$, with $\xi \ll \pi$. Otherwise, the base time-step is too large and most of the electron trajectory has to be sub-cycled. The reduction of $\psi$  at $\xi \ll 1$ occurs due to the denominator in Eq.~(\ref{psi_1}) when the term proportional to $\sin(\xi)$ starts to dominate. In order to find the value of $\xi$ that yields $\psi = \psi_*$, we retain only the $\sin(\xi)$-term in the denominator and expand the expression in Eq.~(\ref{psi_1}) with respect to $\xi$. To the lowest order in $\xi$, we then have
\begin{equation} \label{psi_1_2}
\psi(\xi) = \pi \frac{c \Delta_0}{\lambda} \frac{2}{a_0 \xi^2}.
\end{equation}
It follows from Eq.~(\ref{psi_1_2}) that $\psi = \psi_*$ at 
\begin{equation}
\xi \approx \xi_* \equiv \sqrt{\frac{1}{a_0} \frac{2 \pi}{\psi_*} \frac{c \Delta_0}{\lambda}}.
\end{equation}

The travel time from the stopping point to the point where $\xi = \xi_*$ can be calculated using Eq.~(\ref{Eq_t}). Taking into account that $\gamma = 1 + a^2(\xi)/2$ and assuming that $\xi \ll 1$, we find that to the lowest order in $\xi_*$ this travel time is
\begin{equation} \label{t*}
\frac{c t}{\lambda} = \frac{c t_*}{\lambda} \equiv \frac{1}{2 \pi} \frac{a_0^2}{2} \frac{\xi_*^3}{3}.
\end{equation} 
On the other hand, the travel time to the next stopping point is 
\begin{equation} \label{ts}
\frac{ct}{\lambda} = \frac{ct_s}{\lambda} \equiv \frac{a_0^2}{8}.
\end{equation}
We found this expression by performing the integration in Eq.~(\ref{Eq_t}) from $\xi = 0$ to $\xi = \pi$ and retaining only the leading term, which involves $a_0$. 

The total number of time-steps with $\Delta t = \Delta_0$ between two stopping points is
\begin{equation}
N_0 = t_s / \Delta_0.
\end{equation}
The total number of time-steps $N_*$ with the sub-cycling can be estimated by using $\Delta t = \Delta t_{\min}$ for the time interval that requires the sub-cycling:
\begin{equation}
N_* = \frac{t_s - 2t_*}{\Delta_0} + \frac{2t_*}{\Delta t_{\min}}.
\end{equation}
This gives an upper estimate, since the actual algorithm is adaptive and not all time-steps are as small as  $\Delta t_{\min}$. The relative increase in the total number of time-steps is
\begin{equation}
\frac{N_* - N_0}{N_0} \approx  \frac{2t_* }{ t_s} \frac{\Delta_0}{\Delta t_{\min}}.
\end{equation}
We next use the expressions for $t_*$, $t_s$, $\Delta t_{\min}$, and $\xi_*$ to obtain that
\begin{equation} \label{efficiency}
\frac{N_* - N_0}{N_0} \approx  \frac{2}{3 \pi} a_0^2 \xi_*^5 = \frac{2}{3 \pi} \frac{1}{\sqrt{a_0}} \left(  \frac{2 \pi}{\psi_*} \frac{c \Delta_0}{\lambda} \right)^{5/2}.
\end{equation}
Therefore, the relative number of extra time-steps decreases at least as fast as $1/\sqrt{a_0}$, which makes the sub-cycling mechanism more efficient at higher wave amplitudes.

Finally, it is important to point out the dual role of the adjustable parameter $\psi_*$. On one hand, the number of extra steps increases as $1/\psi_*^{5/2}$ according to Eq.~(\ref{efficiency}), which means that that the efficiency of the sub-cycling decreases with a decrease of the critical rotation angle $\psi_*$. On the other hand, the error in the dephasing also decreases with a decrease of $\psi_*$. This can be illustrated using the results presented in Fig.~\ref{fig:pusher_0s}. In addition to the oscillations in the dephasing between the stopping points, there is also a clear downwards shift introduced by changing the size of the time-step during the sub-cycling procedure. This trend leads to an error of roughly 8\% at the end of the run for $a_0 = 25$. By reducing $\psi_*$ from 0.05 to 0.025 and holding other parameters fixed, we have reduced the dephasing error to less than 2\%. Therefore, the value of the critical rotation angle $\psi_*$ must be carefully chosen such that the desired precision is achieved without sacrificing the efficiency.


\section{Summary and discussion} \label{Sec_6}

We have revisited the classic test problem of a single electron irradiated by a high-intensity plane electromagnetic wave to examine the performance of a particle-in-cell code. We found that the numerical accuracy consistently deteriorates with increasing wave amplitude for a fixed time-step. We have separately examined the accuracy of the standard Boris particle pusher and found that the particle pusher introduces significant errors while integrating the electron motion in the vicinities of stopping points. Field amplitude scans reveal that the deviation of the dephasing rate from unity increases with $a_0$ regardless of the time-step. The errors in the energy gain have a threshold behavior with increasing $a_0$ caused by increased error in the dephasing rate and the threshold amplitude scales inversely proportional to $\Delta t$. 

Based on these observations, we have derived a convenient criterion for the time-step used in the particle pusher:
\begin{equation} \label{condition_final}
c \Delta t / \lambda \ll 1/a_0.
\end{equation}
Our analysis shows that the stopping points are more prone to numerical errors than other parts of the electron trajectory. Numerical errors  from the integration of the electron equation of motion near a stopping point lead to considerable error in the dephasing rate. The error in the momentum gain is relatively small at this stage, however, the error in the dephasing rate affects the subsequent electron acceleration by the wave causing considerable error in the maximum electron energy gain. 

We have developed an efficient algorithm that allows us to reduce the numerical errors generated by the particle pusher. The algorithm uses the derived criterion to adaptively decrease the time-step to ensure that the criterion is always satisfied. We have demonstrated that this adaptive sub-cycling actually becomes more efficient at higher wave amplitudes, so that a significant improvement of the numerical results can be achieved using only a modest increase in the total number of time-steps.

In our analysis, we have used a standard Boris pusher to calculate the electron dynamics, since this is the integrator implemented in some of the PIC codes frequently used to model laser-plasma interactions. It has been recently pointed out that the Boris pusher might lead to errors when calculating orbits of relativistic electrons~\cite{Vay2008}. This can be important on those segments of the electron trajectory where the electron is moving forward with a high $\gamma$-factor and the electric and magnetic field contributions in the equation of motion almost cancel each other out. The Boris pusher does not correctly cancel the electric and magnetic field contributions in the case of a relativistic electron, which can lead to a spurious force~\cite{Vay2008}. The spurious force and the resulting errors can be eliminated by using the Vay particle pusher~\cite{Vay2008} instead of the Boris pusher. In the regime considered in this paper, the relative role of these errors is however not as significant as the role of the errors from integrating the electron equation of motion near the stopping points. This is evident from the dramatic improvement that we have achieved by implementing the sub-cycling algorithm while still using the Boris pusher.

Our discussion has so far been focused on the particle pusher as the main source of numerical errors. The degradation in numerical accuracy observed in PIC simulations under these conditions appears to be primarily from the particle pusher, rather than the field advance. However, another potential source of errors is the numerical dispersion produced by the field solver. It is well known that the frequency $\omega$ of a wave with wave-length $\lambda$ is less than $\omega_* \equiv 2 \pi c/\lambda$:
\begin{equation}
\frac{\omega - \omega_*}{\omega_*} \approx -\frac{\pi^2}{6} \left( \frac{\Delta x}{\lambda} \right)^2 \left[ 1-  \left( \frac{c \Delta t}{\Delta x} \right)^2 \right].
\end{equation}
This error can be greatly reduced by keeping the ratio $c \Delta t/\Delta x$ close to unity. The corresponding discussion can be found in Ref. [\citen{Birdsall2002}].

The spatial grid can also lead to numerical electron heating. The heating is caused by nonphysical instabilities that develop if the grid size exceeds the Debye length~\cite{Birdsall2002c}. The resulting electron energy $\varepsilon$ can be estimated by noting that the heating stops once the Debye length $\lambda_D \approx \sqrt{\varepsilon/4 \pi n e^2}$ becomes comparable to $\Delta x$, where $n$ is the plasma electron density. We are interested in a regime where $n$ is significantly under-dense for a given frequency $\omega$ of the laser pulse irradiating the plasma. It is then convenient to introduce a critical density $n_{\mbox{cr}}$ defined by the condition $\sqrt{4 \pi n_{\mbox{cr}} e^2 / m_e} = \omega$. The wave propagation in a significantly under-dense plasma is similar to that in a vacuum and thus we effectively have $\omega \approx 2 \pi c / \lambda$. Taking this relation and the definition for $n_{\mbox{cr}}$ into account, we find that the condition $\lambda_D \approx \Delta x$ yields
\begin{equation} \label{lambda_D}
\varepsilon \approx m_e c^2 \frac{n}{n_{\mbox{cr}}} \left( 2 \pi \frac{\Delta x}{\lambda} \right)^2.
\end{equation}
Typically, the grid-size is much less than $\lambda / 2 \pi$, so that the electron energy resulting from numerical heating is smaller than $m_e c^2 n/n_{\mbox{cr}}$. This energy is non-relativistic in a significantly under-dense plasma with $n \ll n_{\mbox{cr}}$. On the other hand, the maximum energy of an electron accelerated by a wave with amplitude $a_0$ is $m_e c^2 a_0^2/2$. This energy greatly exceeds $\varepsilon$, provided that $a_0 \gg 1$ and $n \ll n_{\mbox{cr}}$. Moreover, the dephasing rate $I = \gamma - p_x/m_e c$ [see Eq.~(\ref{deph_est})] remains close to unity despite the numerical heating since $\varepsilon \ll m_e c^2$. This means that the electron acceleration by the wave remains unaffected. Therefore, it is not necessary to resolve the Debye length when simulating electron acceleration by an ultra-intense laser ($a_0 \gg 1$) in a significantly under-dense plasma, since the energy that results from numerical heating is inconsequential.

The errors from the particle pusher that we have examined here should be particularly critical when simulating electron acceleration in an underdense plasma. In a two and three-dimensional set-up, only a small group of electrons are accelerated directly by the laser pulse. Our approach to adaptive sub-cycling would be well suited in this case to improve the accuracy of the numerical results. Our algorithm would automatically single out energetic electrons accelerated by a high-amplitude field, leaving the time-step for the other electrons unchanged. In order to implement the adaptive sub-cycling in a given particle-in-cell code, one has to orbit-average the current density of the sub-cycled electrons in a way similar to those discussed in Refs. [\citen{Cohen1985}] and [\citen{Chen2014}]. The orbit-averaged current can then be directly used in the field solver to calculate the fields using the global time-step.


\section{Acknowledgments}

AVA would like to thank Dr. S.P.D. Mangles and Dr. V. N. Khudik for stimulating discussions and constructive comments. Simulations for this paper were performed using the EPOCH code (developed under UK EPSRC grants EP/G054940/1, EP/G055165/1 and EP/G056803/1) using HPC resources provided by the Texas Advanced Computing Center at The University of Texas. AVA was supported by AFOSR Contract No. FA9550-14-1-0045, 
National Nuclear Security Administration Contract No. DE-FC52-08NA28512 and U.S. Department of Energy Contract No. DE-FG02-04ER54742. GEC received support from NNSA Contract No. DE-NA0001976.


\end{document}